\documentclass[9pt,sigconf,nonacm]{acmart}

\usepackage{algorithmic}
\usepackage{graphicx}
\usepackage{textcomp}
\usepackage{xcolor}
\usepackage{enumitem}

\def\BibTeX{{\rm B\kern-.05em{\sc i\kern-.025em b}\kern-.08em
    T\kern-.1667em\lower.7ex\hbox{E}\kern-.125emX}}

\usepackage{amsmath}
\usepackage{algorithmic}
\usepackage{array}
\usepackage{comment}
\usepackage{listings}
\usepackage{graphicx}

\usepackage[font=small]{caption}

\begin{document}

\title{Understanding Inference-Time Token Allocation and Coverage Limits in Agentic Hardware Verification}

\author{Vihaan Patel, Vidya Chhabria, Aman Arora}
\affiliation{%
  \institution{Arizona State University}
  \city{Tempe}
  \state{Arizona}
  \country{USA}
  }
\email{{vpatel69, vachhabr, aman.kbm}@asu.edu}

\begin{abstract}
Coverage closure is the most time-consuming phase of hardware verification, and recent large language model (LLM)-based coding agents offer a promising approach to automated stimulus generation. However, prior LLM-based flows do not systematically analyze which coverage holes remain difficult to close or how inference-time computation is allocated during agentic verification. As a result, the efficiency limits and failure modes of LLM-based coverage closure remain poorly understood. We present an empirical study using a two-tier agentic framework comprising a base  agent and an enhanced domain-specialized  system.
We instrument the framework to track token usage across six categories, including system prompt, design comprehension, stimulus generation, coverage feedback, error recovery, and agentic overhead. 
We show that domain specialization shifts token allocation toward coverage-directed reasoning and improves efficiency. 
Our framework enables a taxonomy of coverage holes:  methodology-bound ceilings (integration tied-off hardware, infeasible boundaries, dead code) and  reasoning frontiers (protocol sequencing, multi-module pipeline warm-up, narrow timing conditions), exposing fundamental limits of LLM-driven approaches.  
Across designs, the enhanced system achieves comparable or higher coverage (up to 100\%) while using 4–13$\times$ fewer tokens and converging to coverage targets 2–4$\times$ faster than a general-purpose baseline.
Our results characterize the limits of LLM-based coverage closure and guide profiling-driven efficient agent design for hardware verification.
\end{abstract}
\vspace{-5mm}
\maketitle

\section{Introduction} \label{sec:intro}

Hardware verification consumes an estimated 60--70\% of chip design effort \cite{foster:2024}, often dominating engineering cost and schedule. Within verification, coverage closure remains one of the most time-consuming and iterative stages. 
Verification engineers manually interpret specifications, draft testplans, generate stimulus, analyze coverage reports, and refine their approach across dozens of iterations, which can take weeks to months for complex designs.

Recent advances in LLMs have enabled agentic coding systems that generate non-trivial programs and interact with toolchains. Prior work has explored LLM-assisted RTL design generation \cite{blocklove:chipchat:2023, allam:rtlrepo:2024, delorenzo:creativeval:2024, liu:rtlcoder:2024, liu:verilogeval:2023, lu:rtllm:2024a, thakur:autochip:2024, thakur:verigen:2024, wang:chatcpu:2024, zhang:mgverilog:2024, fu2025gpt4aigchipnextgenerationaiaccelerator, hu2024uvllmautomateduniversalrtl}, as well as verification code generation like assertions \cite{mali:chiraag:2024a, fang:assertllm:2024a, qiu:autobench:2024a}.
More recently, LLM-assisted verification flows have shown iterative generation of stimulus \cite{zhang:llm4dv:2023, ma:verilogreader:2024a, hassan:promptverifyrepeat:2025, ye2025conceptpracticeautomatedllmaided}. However, two key questions remain underexplored. \textit{(1) Which coverage holes remain fundamentally difficult to close with LLM-generated stimulus?}
Automatically analyzing coverage holes enables exclusion list generation, informs benchmark design, guides human escalation decisions, and provides actionable feedback to designers and verification engineers (e.g., updating the coverage model or the specification).
 \textit{(2) Where are inference-time tokens allocated during agentic verification?}
Token allocation analysis, similar to profiling, can guide efficient agent design, including multi-agent pipelines, dynamic model selection, and structured tool interfaces, that reduce token usage and ever-increasing agentic deployment costs. 

\begin{figure}[t]
    \centering
    \includegraphics[width=\linewidth]{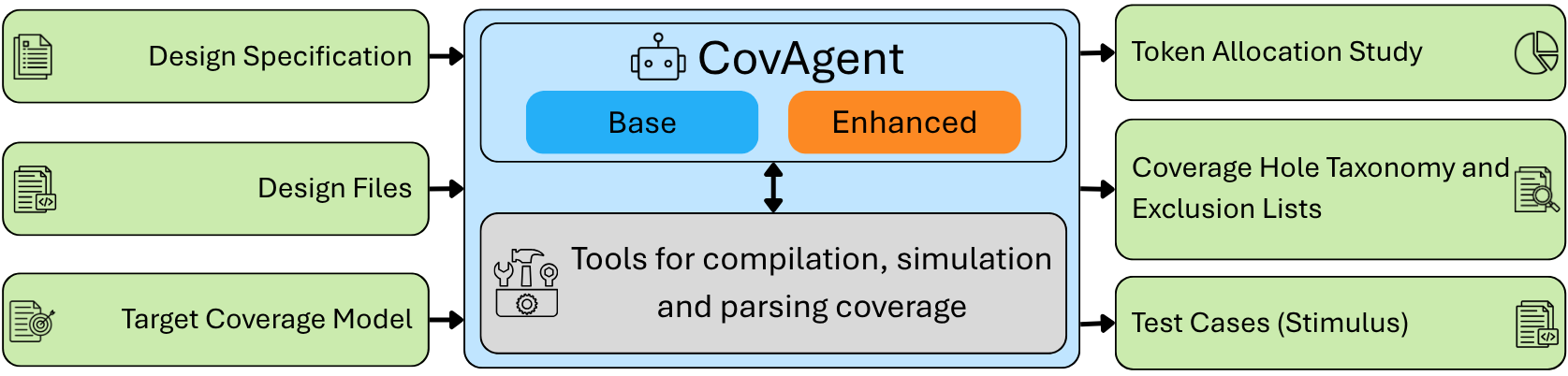}
    \vspace{-5mm}
    \caption{An overview of the proposed framework}
    \label{fig:framework_overview}
\end{figure}

To study these questions, we develop CovAgent, a two-tier agentic framework for coverage closure and analyze whether domain-specialized workflows shift token allocation toward more productive, coverage-directed work.
The \textbf{base agent} is built on a general-purpose coding agent 
using its native feedback \& reasoning loop and enabling interaction with verification tools via broad shell access. 
The \textbf{enhanced agent} is a domain-specialized system 
with structured tool interfaces and coverage-focused feedback.
Fig. \ref{fig:framework_overview} provides an overview of CovAgent. It consumes design specifications, design files, and a target coverage model. 
It generates testcases with the goal of achieving 100\% coverage, a coverage hole classification report (and exclusion lists for human review), and an inference-time token allocation report.
As agentic workflows become increasingly expensive and subject to token limits,  CovAgent's novel instrumentation methodology
enables optimization of token usage, enhancing viability at scale. 
Our contributions are:

\begin{enumerate}[nosep, leftmargin=*]
\item \textbf{CovAgent}: A two-tiered agentic framework, available at \cite{covagent}, for coverage closure comprising a base agent and an enhanced domain-specialized agent, instrumented for collecting token allocation data, enabling comparison of general-purpose vs.\ domain-adapted agentic verification.

\item \textbf{Token breakdown}: 
A six-category breakdown of token allocation during agentic verification, showing that design comprehension reasoning scales disproportionately with design size and only 12--18\% of accumulated tokens produce testbench code.

\item \textbf{Coverage hole taxonomy}: A six-category classification of residual coverage holes that reveals fundamental limits of agentic stimulus generation. The taxonomy is manually analyzed by a verification expert and verified to confirm the cause of the missed coverage and the absence of inaccuracies.

\item \textbf{Metrics for agent efficiency}: New metrics to evaluate agentic verification flows, specifically Coverage-Per-Token (CPT) and Tokens-To-Coverage-Target\_N (TTCT\_N), enabling quantitative comparisons and optimizations.  
\item \textbf{Empirical evaluation}: Experiments on 19 benchmark designs revealing that (a) domain-specialized agent achieves 4--13$\times$ lower token usage with equal or higher coverage (95--99\%) than the base agent, and (b) a smaller model achieves similar results on simpler designs but suffers up to 29pp coverage degradation on complex designs, indicating that the framework offsets reduced model capability only below a complexity threshold.
\end{enumerate}


\section{Related Work} \label{sec:related_work}

\textbf{LLMs for Hardware Design and Verification:}
LLMs have recently been used for RTL generation and debugging~\cite{blocklove:chipchat:2023, allam:rtlrepo:2024, delorenzo:creativeval:2024, liu:rtlcoder:2024, liu:verilogeval:2023, lu:rtllm:2024a, thakur:autochip:2024, thakur:verigen:2024, wang:chatcpu:2024, zhang:mgverilog:2024, fu2025gpt4aigchipnextgenerationaiaccelerator}.
They have also been applied to hardware verification tasks, including assertion synthesis, testbench construction, UVM environment synthesis, and coverage-guided stimulus generation.
UVLLM~\cite{hu2024uvllmautomateduniversalrtl} identifies syntactic and functional RTL errors using an agentic pipeline, while ChiRAAG~\cite{mali:chiraag:2024a} generates assertions via interactive prompting. AssertLLM~\cite{fang:assertllm:2024a} uses ChatGPT and retrieval-augmented generation to produce SystemVerilog assertions from specifications. AutoBench~\cite{qiu:autobench:2024a} generates hybrid testbenches using an HDLBits-derived dataset. LLM4DV~\cite{zhang:llm4dv:2023} generates stimulus from coverpoints across three designs. Verilog Reader~\cite{ma:verilogreader:2024a} focuses on code-coverage-driven testcase generation. Hassan et al.~\cite{hassan:promptverifyrepeat:2025} propose an iterative verification framework where LLMs play a semantic role. Finally, Ye et al.~\cite{ye2025conceptpracticeautomatedllmaided} propose an agentic UVM-based framework with a nine-design benchmark suite.
While these efforts demonstrate feasibility, prior work leaves two gaps which CovAgent aims to fill:
(1) failure to distinguish whether residual coverage holes arise from structural methodology limits or LLM reasoning failures when 100\% coverage is not achieved; and
(2) lack of analysis of inference-time token allocation during agentic verification despite prior research~\cite{snell2024scaling, wu2025inference} showing that token allocation significantly affects task performance.


\noindent\textbf{Agentic Coding Systems:}
General-purpose coding agents such as Codex CLI \cite{codex} and other ReAct-style \cite{react} systems demonstrate tool-augmented reasoning and iterative program repair, combining LLM reasoning with tool execution-based feedback.
However, such agents are not optimized for verification workflows and often spend tokens on environmental exploration, redundant parsing, or unstructured debugging. 
To the best of our knowledge, no prior verification frameworks have combined Codex and LangGraph \cite{langgraph} while analyzing token allocation during verification.

\section{The CovAgent Framework} \label{sec:proposal}

\subsection{Architecture Overview}

CovAgent is an agentic framework designed to systematically study, and ultimately improve, the process of coverage closure in hardware verification. 
Rather than treating LLM-based verification as a monolithic workflow, CovAgent provides two tiers: a base tier using general-purpose OpenAI Codex CLI \cite{codex}, and an enhanced tier using a domain-specialized agent implemented in LangGraph \cite{langgraph}.  
This two-tier design enables controlled experimentation on how architectural constraints, tool abstractions, and feedback structure influence both coverage outcomes and inference-time efficiency.
At its core, CovAgent implements coverage closure as an iterative reasoning loop over six stages: (1) design comprehension from specification and top-level interface, (2) testbench synthesis constrained to stimulus generation, (3) compilation with coverage instrumentation, (4) multi-seed simulation and coverage aggregation, (5) structured analysis of coverage holes, and (6) targeted refinement of stimulus to address those holes. This loop mimics human-based feedback-driven coverage closure.
\textit{Note that our framework is not tied to Codex and can integrate other agents (e.g. Claude Code) with minimal changes.}

\begin{figure}[t]
    \centering
    \includegraphics[width=0.9\linewidth]{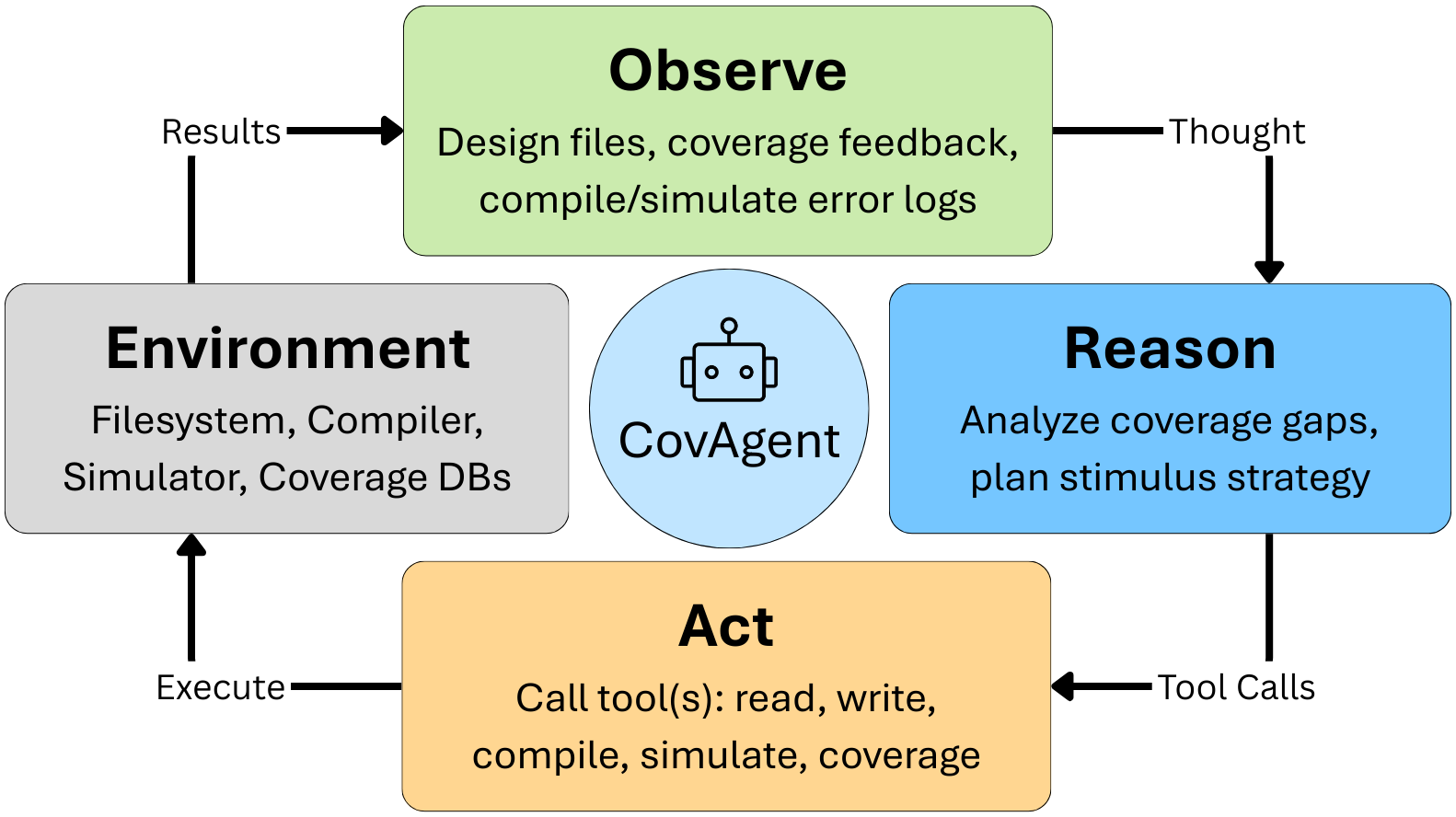}
    \vspace{-2mm}
    \caption{CovAgent workflow}
    \label{fig:covagent_arch}
    \vspace{-7mm}
\end{figure}

As shown in Fig. \ref{fig:covagent_arch}, CovAgent adopts the ReAct (Reasoning + Acting) \cite{react} paradigm. Unlike generic ReAct-based systems, CovAgent augments the reasoning loop with verification-specific context that embeds established hardware verification methodology. The agent's role is defined as an expert verification engineer and it is provided access to the specification, design sources, and the top-level module interface. Strict stimulus-generation constraints (i.e., prohibiting hierarchical force/release and submodule instantiation) are applied ensuring that exploration remains limited to realistically reachable behaviors. In addition, testbench requirements (e.g., signal declaration rules and constrained randomization) together with coverage improvement strategies are provided to the agent.
The coverage-directed stimulus generation loop terminates when any of these conditions are met: 100\% coverage achieved, no coverage progress limit (5 iterations), or context window exhaustion. 
Upon termination, the latest coverage snapshot is saved and the agent writes a structured report.
The agent is instructed to \textbf{reason about why coverage points remain uncovered}, classifying them as unreachable from top-level ports, excludable (dead code), potential bugs, or needs-more-effort. 
This enables generation of coverage exclusion lists and providing 
actionable feedback to designers and verification engineers (e.g., to update coverage models).

To support quantitative analysis, CovAgent incorporates \textbf{fine-grained token-level instrumentation}. 
Every agent interaction is logged and classified into six categories:
System Prompt (input), Design Comprehension (DC) (input + reasoning + output), Stimulus Generation (SG) (reasoning + output), Coverage Feedback (input), Error Recovery, and Agentic Overhead.
As we later show, this profiling reveals that efficiency gains arise not merely from better models, but from restructuring how reasoning effort is spent.

\subsection{Base Agent}


The base agent represents a minimally constrained instantiation of agentic verification. It uses out-of-the-box OpenAI's Codex CLI and its native ReAct style workflow.
It executes the verification workflow by providing the agent with command templates for compilation, simulation, and coverage report generation, in addition to information in the previous sub-section.
This agent maximizes flexibility by providing broad access to a sandboxed execution environment.
It can execute arbitrary commands, write and run auxiliary scripts, and dynamically adjust its workflow. It can batch-generate testbenches, build custom coverage parsers, implement feature-based stimulus generation strategies, or develop iterative debugging pipelines. 


However, this flexibility comes at a cost. The agent must implicitly infer the structure of the verification workflow, and perform environmental exploration, parsing, and adhoc orchestration. 
The absence of structured feedback forces the agent to reconstruct high-level actions (e.g., prioritize coverage holes) from low-level artifacts such as raw reports. As a result, the base agent serves as a reference point for unconstrained agentic verification, capturing the upper bound of generality but also exposing its inefficiencies.

\subsection{Enhanced Agent}

The enhanced tier encodes domain knowledge into the agent through:
%
\textbf{(a) Structured Tool Interfaces:} CovAgent exposes seven tools: read\_file, write\_file, list\_directory for filesystem access; compile\_\\
design and run\_simulation for simulator interaction; parse\_coverage for coverage analysis; and run\_verification\_cycle, a composite tool that chains write, compile, simulate, and parse into a single atomic call. 
By collapsing multiple steps into a single tool call, the composite tool reduces unnecessary deliberation and minimizes token overhead associated with intermediate reasoning.
%
%
%
%
\textbf{(b) Structured Coverage Feedback:} The system provides curated, prioritized information about uncovered coverage points. This transforms coverage analysis from a low-level parsing task into a high-level decision problem, allowing the agent to focus its reasoning on what to do next.
\textbf{(c) LangGraph State Graph:} The enhanced agent implements the ReAct loop as a state graph with six nodes: Initialize (loads configuration, creates work directories, builds system prompt), Agent (invokes LLM with tools), Tools (LangGraph's built-in ToolNode executes tool calls), Update State (parses tool results, updates coverage and failure trackers), Prune Context (removes old error logs) and Finalize (injects termination message instructing the agent to write its report).

 \begin{figure}[t]
     \centering
     \includegraphics[width=\linewidth]{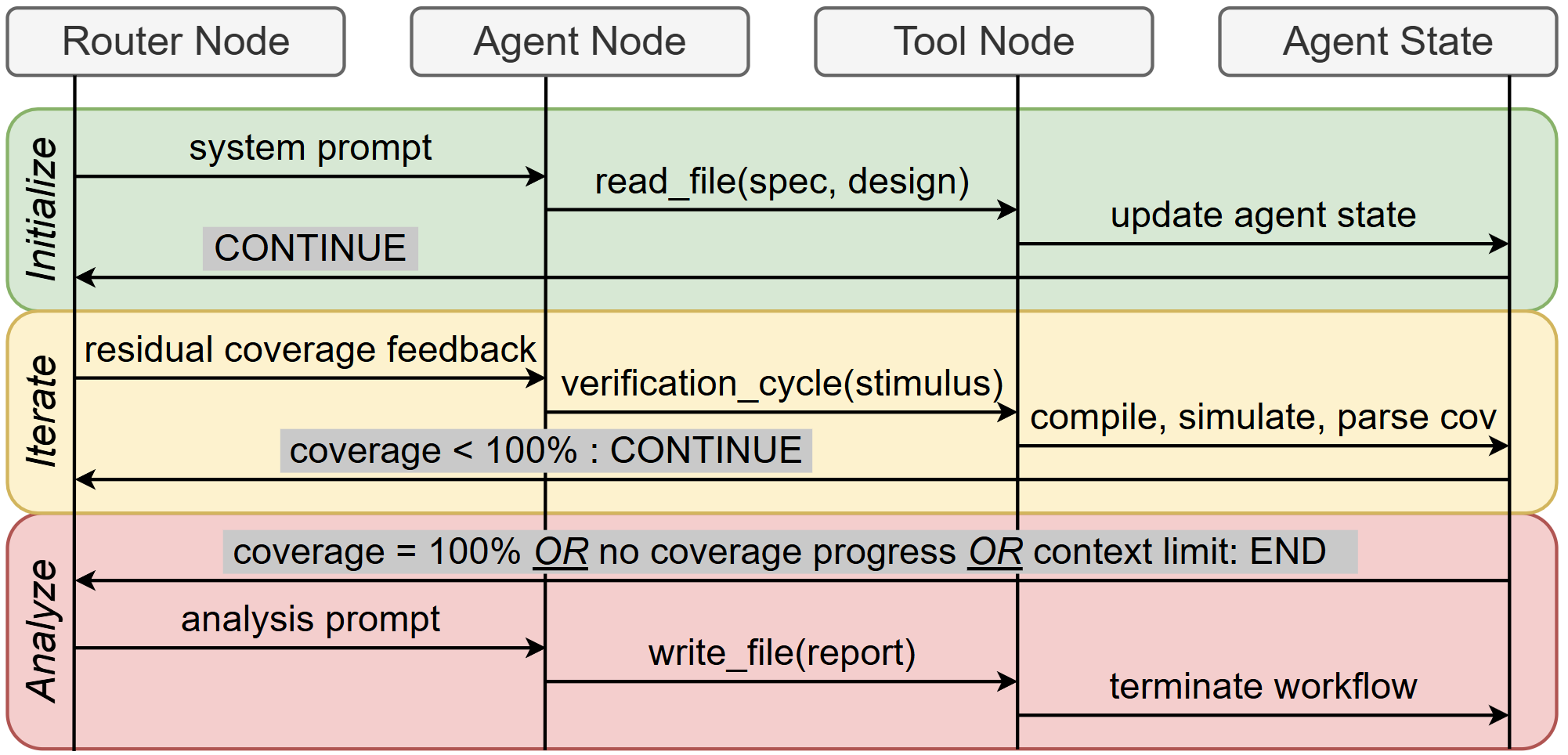}
     \caption{Architecture of the enhanced agent}
     \label{fig:enhanced_arch}
     \vspace{-6mm}
 \end{figure}

Fig. \ref{fig:enhanced_arch} shows the architecture of this enhanced agent. During initialization, the Router delivers the system prompt and the Agent is given access to the specification and design files via read\_file, with Update State recording the context consumed. The workflow then enters an iterative loop: the Agent calls run\_verification\_cycle with its generated stimulus, the Tool Node executes the composite write, compile, simulate, and parse pipeline atomically, and Update State merges the new coverage database and checks termination conditions. If no termination condition is met, the Router injects structured coverage feedback and returns control to the Agent for a refined iteration. When termination conditions are met, the Router triggers the analysis phase: a finalize prompt instructs the Agent to write a report classifying residual coverage.

\begin{table}[t]
\centering
\setlength{\tabcolsep}{3pt}
\caption{Designs used for our study}
\vspace{-3mm}
\small
\begin{tabular}{|p{0.4cm} | p{3cm} | p{1cm}|  p{1cm}| p{2cm}|}
\hline
\textbf{Key} & \textbf{Design} & \textbf{Hier. Lvls} & \textbf{Line Count} & \textbf{FSMs \& States} \\
\hline
e1 & ttc\_lite~\cite{pinckney2025comprehensiveverilogdesignproblems}                    & 1  & 101  & 0 \\
e2 & memory\_scheduler~\cite{pinckney2025comprehensiveverilogdesignproblems}            & 1  & 129  & 0  \\
e3 & door\_lock~\cite{pinckney2025comprehensiveverilogdesignproblems}                   & 1  & 137  & 1 [7] \\
e4 & sorter~\cite{pinckney2025comprehensiveverilogdesignproblems}                       & 1  & 164  & 1 [7] \\
e5 & float\_multiplier~\cite{samidhm_float}                                             & 2 & 218   & 0  \\
e6 & poly\_interpolator~\cite{pinckney2025comprehensiveverilogdesignproblems}           & 2  & 224  & 0  \\
e7 & rgb\_color\_space\_conv~\cite{pinckney2025comprehensiveverilogdesignproblems} & 1 & 282    & 0  \\
e8 & spi\_complex\_mult~\cite{pinckney2025comprehensiveverilogdesignproblems}           & 2  & 348  & 1 [4]  \\
e9 & pooling~\cite{utlca_pooling}                                     & 1  & 412   & 0  \\
e10 & cryptech\_uart~\cite{secworks_uart}            & 2  & 447   & 2 [5,5] \\
e11 & float\_adder~\cite{samidhm_float}                                & 2 & 463  & 0   \\
e12 & sha1\_top~\cite{secworks_sha1}                 & 3  & 630   & 1  [3] \\
e13 & chacha\_top~\cite{secworks_chacha}             & 3  & 778   & 1  [8] \\
h1 & sd\_controller\_top~\cite{mczerski_sdcard}     & 5  & 2154   & 5 [3,8,4,6,3] \\
h2 & trng\_top~\cite{secworks_trng}                 & 4 & 3602   & 7  [10,3,2,7,6,8,4] \\
h3 & can~\cite{can_design}                 & 3 & 3722   & 5  [16,5,3,3,3] \\
h4 & axi\_crossbar~\cite{axi_crossbar}       & 5 & 4613   & 0   \\
h5 & jpeg\_core~\cite{jpeg_core}     & 4  & 6232   & 4   [18,6,3,3]\\
h6 & ethernet\_mac~\cite{eth_mac}     & 5  & 8580   & 4  [10,6,3,3] \\

\hline
\end{tabular}
\label{tab:design-difficulty}
\vspace{-4mm}
\end{table}

\section{Methodology} \label{sec:methodology}


\noindent
\textbf{Models:}
All experiments use GPT-5.2 as the primary LLM, used via the OpenAI API.
GPT-5-mini is used for a sensitivity study with the enhanced agent to understand the impact of model capability.
\textit{Note that our framework supports open-weight models as well. We experimented with Llama and Qwen models, but performance lagged behind commercial models, leading us to use GPT models. Newer open-source models may achieve comparable results.}

\noindent\textbf{Metrics:}
We evaluate performance using multiple metrics.
\textit{Final coverage (\%)}: Cumulative coverage after all iterations. 
\textit{Total token consumption}: Sum of input,  output, and reasoning tokens across all API calls.
\textit{Token allocation by category}: Post-run classification into six categories (Section \ref{sec:proposal}). 
\textit{Cost}: The dollar cost of all tokens calculated based on current pricing.
\textit{Coverage-Per-Token (CPT)}: Coverage percentage gained per 10K tokens.
\textit{Tokens-To-Coverage-Target\_N (TTCT\_N)}: Number of total tokens used to achieve N\% coverage score. In this work, we use N=100\%. Other values of N can be used to obtain further insights. 

\noindent\textbf{Runs and Seeds:}
All results are averaged over 3 runs to account for LLM stochasticity. The achieved coverage varied 0-2\% across runs.
Each testbench iteration uses 5 random seeds, with coverage merged per-iteration and cumulatively, reflecting standard industry verification methodology.

\noindent
\textbf{Benchmarks:}
We use a set of 19 hardware designs (Table \ref{tab:design-difficulty}) 
sourced 
from the CVDP benchmark suite \cite{pinckney2025comprehensiveverilogdesignproblems} and GitHub.
CVDP designs are from the verification problems (cid12-14) in the suite. Using only problems with complete design specifications and code resulted in 14 designs. 7 were further removed because they were too small (under 100 lines of code (LOC)) and not sufficiently challenging for modern LLMs. Harder and larger GitHub designs were added to create a more challenging benchmark set than CVDP.
The designs span timer/counter, memory, FSM, arithmetic datapath, cryptography, communication, protocol controller, image proc, and network peripherals.
We categorize these designs into two complexity levels: easy and hard, based on multiple features: LOC, module hierarchy levels, number of finite state machines (FSMs), and number of states per FSM.   
We normalize each feature to the maximum value of the feature across designs resulting in each feature's value between 0 and 1. Then, we perform Principal Component Analysis (PCA). 
The first two principal components explained over 90\% of the variance. PCA visualization revealed two natural groups or categories, which we label ``easy'' (13 designs) and ``hard'' (6 designs).

Note that these designs are representative of unit level verification in the industry. Larger designs are verified at sub-system or top level and primarily involve connectivity testing. Consequently, randomized stimulus generation and full coverage are less critical at these levels. While our evaluation focuses on unit level verification, our agent is equally applicable to sub-system and top-level environments - only the coverage model will change. We leave this as future work. Nevertheless, our benchmarks are larger and more complex than those used in prior work \cite{qiu:autobench:2024a, zhang:llm4dv:2023, ma:verilogreader:2024a, ye2025conceptpracticeautomatedllmaided}.

\vspace{-3mm}
\section{Results} \label{sec:results}

\subsection{Coverage Curves}

\begin{figure}[t]
    \centering
    \includegraphics[width=\linewidth]{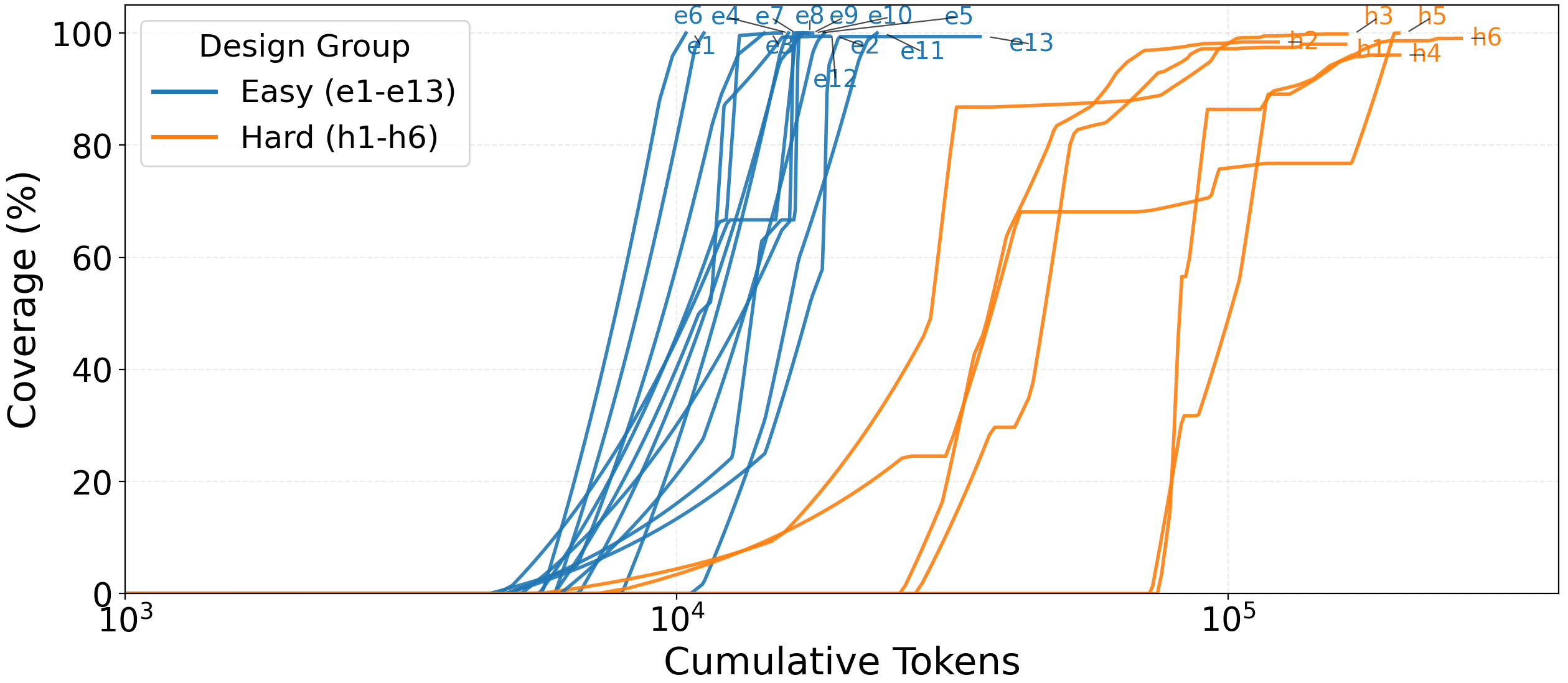}
    \vspace{-7mm}
    \caption{Coverage achieved vs. cumulative tokens across all designs}
    \label{fig:coverage_curves}
    \vspace{-5mm}
\end{figure}

Fig. \ref{fig:coverage_curves} shows coverage achieved by our enhanced framework for the various designs used in our study, and analyze it as a function of cumulative tokens (context window), a continuous view of the Coverage-Per-Token (CPT) metric. 

\textbf{Convergence Pattern.}
Easy designs converge to 100\% within ${\sim}$10K–30K tokens with a steep, near-vertical rise. Hard designs require ${\sim}$55K–175K tokens and plateau at 96–99\%. 
All curves begin with a 0\% coverage region corresponding to tokens spent before the first testbench compiles and simulates. For easy designs, this is ${\sim}$5K–10K tokens; for hard designs, ${\sim}$25K–75K tokens. 

\textbf{Diminishing Returns and Last-Mile Effort}
Across all designs, we observe a consistent diminishing-returns pattern. Most designs exhibit a steep initial coverage slope, rapidly approaching a design-specific plateau (${\sim}95\%$), after which the marginal utility of additional tokens drops sharply. A significant portion of the tokens is utilized after this milestone, producing only incremental gains. This mirrors human hardware verification flows where typically, the last 5\% of coverage requires the most effort.

\textbf{Reasoning Effort and Design Complexity.}
Scaling trends further reveal that reasoning effort grows disproportionately with design complexity. 
Harder designs require additional reasoning for design comprehension before effective stimulus generation.
When normalized by LOC, reasoning tokens per line increase dramatically with scale: easy designs require approximately 0.1–0.3 reasoning tokens per LOC, whereas hard designs require approximately 4–8 reasoning tokens per LOC.
This behavior highlights both the strengths and the fundamental limits of scaling LLM-driven coverage closure.

\subsection{Inference-Time Token Allocation}

\begin{figure}[t]
    \centering
    \includegraphics[width=\linewidth]{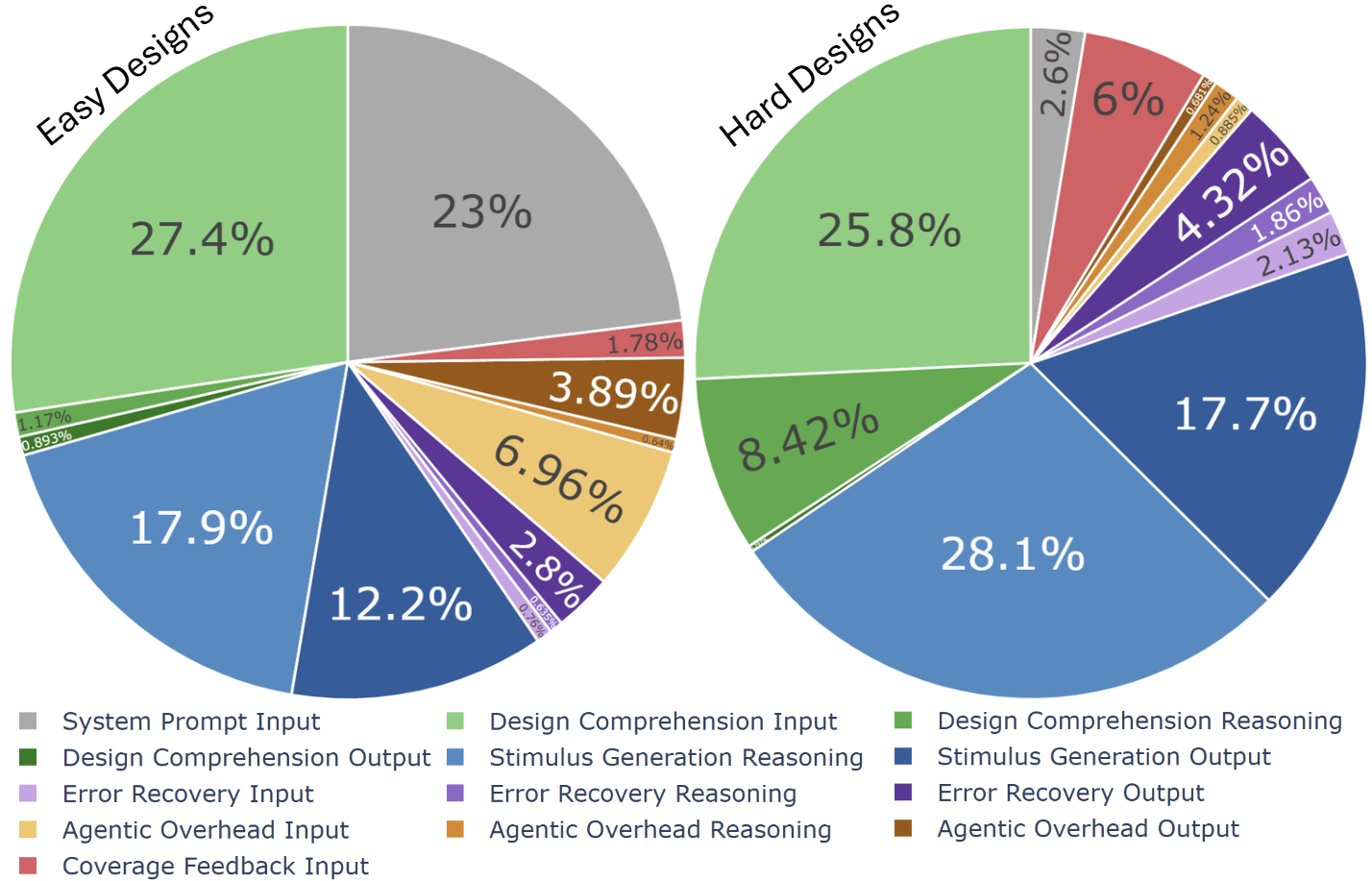}
    \caption{Average token allocation across designs}
    \label{fig:token_alloc}
    \vspace{-4mm}
\end{figure}

Fig.~\ref{fig:token_alloc} shows token allocations averaged across easy and hard designs generated by CovAgent. 

\textbf{Effective Artifact Ratio.}
Only $12$--$18\%$ of accumulated tokens correspond to SG output - the actual testbench code. The remaining tokens are spent on comprehension, reasoning, feedback, and infrastructure. Iterative coverage refinement is therefore token-intensive where most tokens support the loop rather than produce code. 

\textbf{Allocation Shifts with Scale and Complexity.}
For easy designs, the System Prompt ($23\%$), DC input ($27\%$), and SG ($30\%$) dominate. Coverage Feedback and Error Recovery remain negligible ($<5\%$). The agent typically reads the design, generates a testbench, and converges within one or two iterations.
For hard designs, the profile shifts substantially. The System Prompt share drops to $2.6\%$ (amortized with increasing context). SG reasoning increases ($18\% \rightarrow 28\%$), DC reasoning (${\sim}1\% \rightarrow 8\%$), and Coverage Feedback ($6\%$) appear. Error Recovery rises to $8\%$, reflecting the increased compile and simulation failures that accompany complex designs. 
Notably, DC input remains ${\sim}26\%$ across complexities. The model reads a similar proportion of context but the hidden reasoning cost within DC expands $8\times$.

\textbf{Reasoning Redistribution and Plateau Effects.}
Examining reasoning tokens alone clarifies the scaling behavior. In easy designs, $89\%$ of reasoning tokens are devoted to SG; the model quickly understands the design and focuses its thinking on code generation. In hard designs, DC absorbs $21\%$ of the reasoning budget, forcing the model to split effort between understanding the design and generating stimulus. For hard designs, 24\% of DC tokens are reasoning, compared to 4\% for easy designs. This indicates that design comprehension is the bottleneck across the hard designs. The SG reasoning to SG output ratio remains relatively stable ($\sim1.5\times$), suggesting that code-generation deliberation scales modestly; it is comprehension effort that explodes.


\subsection{Coverage Hole Taxonomy}

For each run, CovAgent writes a post-run report analyzing remaining uncovered coverage points. We classify every coverage hole from the 6 sub-100\% designs into 6 categories, tagged as methodology-bound (ceiling) or reasoning-failure (frontier).
Table \ref{tab:coverage_taxonomy} shows this taxonomy.
The taxonomy was manually audited by a verification expert by analyzing each coverage hole and verifying why it was missed. Future work involves adding an ``LLM-as-a-judge'' validation pass that generates reports for human audit.

\renewcommand{\arraystretch}{0.8}
\setlength{\tabcolsep}{3pt}

\begin{table}[t]
\centering
\small
\setlength{\tabcolsep}{2pt}
\renewcommand{\arraystretch}{0.9}
\begin{tabular}{|>{\centering\arraybackslash}c|p{1.8cm}|l|p{3.7cm}|c|}
\hline
\textbf{ID} & \textbf{Category} & \textbf{Tag} & \textbf{Description} & \textbf{\% Count} \\
\hline
M1 & Integration Tied-Off Hardware & Ceiling & Internal signals hardwired to constants at integration level & 4.1 \\
\hline
M2 & Infeasible Boundaries & Ceiling & Coverage points requiring astronomically many simulation cycles & 0.01 \\
\hline
M3 & Defensive/ Dead Code & Ceiling & FSM defaults, debug paths, redundant conditions & 2.7 \\
\hline
R1 & Protocol Sequencing Complexity & Frontier & Multi-step protocol handshakes requiring correct bus functional models & 40.2 \\
\hline
R2 & Multi-Module Pipeline Warm-up & Frontier & Deep pipeline chains requiring coordinated activation & 49.9 \\
\hline
R3 & Narrow Timing \& Rare Input & Frontier & Cycle-precise alignment or sparse numerical conditions & 2.3 \\
\hline
\end{tabular}
\caption{Classification of coverage holes into methodology ceilings (M) and reasoning failures (R). \% Count shows the percentage of total coverage holes seen across designs for each category.}
\label{tab:coverage_taxonomy}
\vspace{-10mm}
\end{table}



\textbf{Reasoning failures dominate on complex designs.}
While aggregate numbers show a roughly even split (38 ceiling vs. 35 frontier), the distribution is sharply skewed by complexity. \textit{Chacha} has only methodology-bound holes. The complex designs (\textit{ethmac}, \textit{SD-card}, \textit{trng}, \textit{CAN}, \textit{AXI crossbar}) account for all 35 reasoning-failure holes, concentrated in protocol sequencing (R1). \textit{Ethmac} is particularly interesting: 80\% of its residual coverage points are reasoning failures where the agent needed to construct Wishbone burst models, MII frame generators, and MDIO PHY responders. \textit{CAN} controller reinforces this pattern: requiring bit-accurate CAN frame generation, bit stuffing, and BTL-aligned timing.
While the agent correctly diagnoses these problems, it fails to implement the solutions. This gap between diagnostic and generative capabilities is a key finding of our framework.

\textbf{Methodology-bound holes define the achievable ceiling.}
These holes cannot be closed without environmental changes such as exposing tied-off registers, adding Design-For-Test (DFT) hooks, or formal reachability analysis. 
They account for a small portion (<7\%) of the total residual coverage holes in our benchmarks. 


The coverage hole taxonomy maps directly to industry coverage exclusion and waiver practices. The methodology-bound categories (M1–M3) are exclusion targets; the reasoning frontiers (R1–R3) identify where human escalation (e.g., for spec updates) or Bus Functional Model (BFM) development would be useful. 
CovAgent generates coverage exclusion lists that can be reviewed by  human engineers and incorporated into the verification  environment.

\subsection{Impact of Domain Specialization}

\begin{figure}[t]
    \centering
    \includegraphics[width=\linewidth]{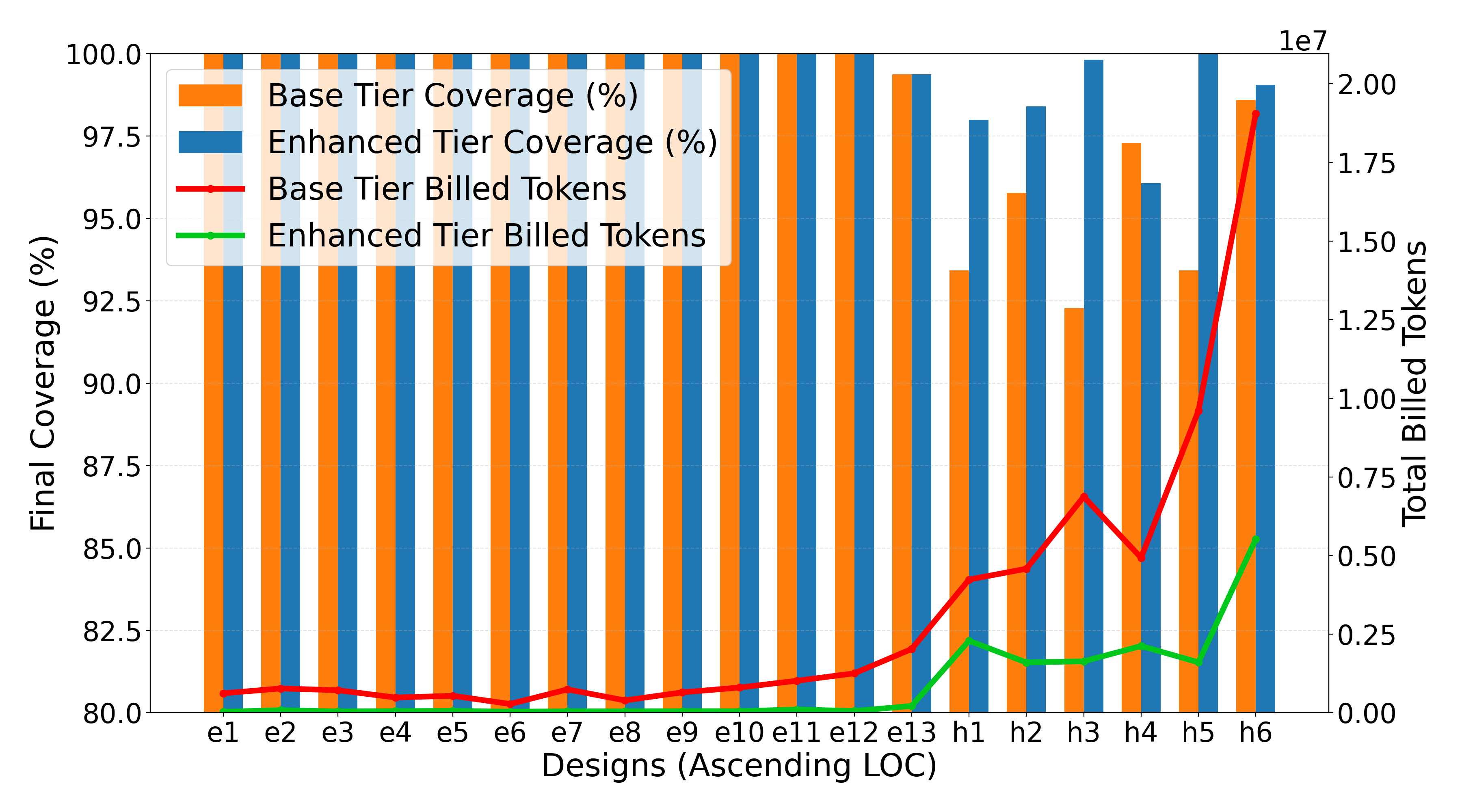}
    \caption{Comparison between CovAgent's base and enhanced tiers}
    \label{fig:domain_specialization_impact}
    \vspace{-6mm}
\end{figure}

Fig. \ref{fig:domain_specialization_impact} and Table \ref{tab:ratio-metrics} compare the results obtained by using CovAgent's base tier and the enhanced domain-specialized tier. 



\textbf{Coverage parity at order-of-magnitude lower cost.}
On 12 easy designs both tiers reach 100\% coverage, but the enhanced tier achieves $0.04$--$0.12$ lower TTCT\_100 because of higher token usage in the base tier resulting from environment and parsing overhead. Other designs do not reach coverage closure in either tier, so TTCT\_100 is not applicable.

\textbf{On hard designs the advantage compresses.}
The enhanced tier reaches equal or higher coverage on 4 hard designs ($+0.5$ to $+7.5$pp) due to effective feedback that directs the agent toward uncovered coverage points. 
CPT stays above 1 on most, but nears parity on \textit{trng} ($1.07$) and \textit{ethmac} ($1.02$). 
The two exceptions are \textit{SD-card} ($0.45$) and \textit{AXI crossbar} ($0.44$). 
\textit{SD-card} is an anomaly where the enhanced tier achieves lower coverage than the base tier (more runs may result in higher coverage), while \textit{AXI crossbar} exhibits the opposite behavior with substantially higher coverage. These results indicate that CPT alone is insufficient for comparing agent efficiency when final coverage levels differ. Since CPT is computed at each agent's stopping point, it does not account for differences in achieved coverage - hence the need for both CPT and TTCT\_N.

\textbf{The token gap is driven by infrastructure overhead.}
Across designs, the base tier utilizes 400K--1.2M total billed tokens, while the enhanced domain-specialized agent spends 30K--100K. This confirms the overhead is dominated by general-purpose environment interaction, not design-specific reasoning.
The enhanced agent shifts allocation toward reasoning while structured coverage feedback reduces wasted tokens.
Thus, domain specialization reduces unproductive exploration and accelerates convergence. 
This paves way for designing custom domain-specialized pipelines, with potentially using smaller LLMs to reduce costs and improve efficiency.

\begin{table}[t]
\centering
\setlength{\tabcolsep}{3pt}
\caption{Comparing CovAgent's base and enhanced tiers using CPT and TTCT\_100. CPT > 1 indicates the enhanced agent provides higher coverage per 10k tokens consumed. TTCT\_100 < 1 indicates the enhanced agent needs fewer tokens to achieve 100\% coverage. N/A is used for TTCT\_100 when the design did not achieve 100\% coverage.}
\vspace{-2mm}
\small
\begin{tabular}{|p{0.4cm} | p{3cm} | p{1cm}|  p{1.5cm}|}
\hline
\textbf{Key} & \textbf{Design} & \textbf{CPT} & \textbf{TTCT\_100} \\
\hline
e1 & ttc\_lite~\cite{pinckney2025comprehensiveverilogdesignproblems}                    & 2.85  & 0.05 \\
e2 & memory\_scheduler~\cite{pinckney2025comprehensiveverilogdesignproblems}            & 1.82  & 0.10  \\
e3 & door\_lock~\cite{pinckney2025comprehensiveverilogdesignproblems}                   & 1.77  & 0.07  \\
e4 & sorter~\cite{pinckney2025comprehensiveverilogdesignproblems}                       & 2.89  & 0.11  \\
e5 & float\_multiplier~\cite{samidhm_float}                                             & 1.79 & 0.11   \\
e6 & poly\_interpolator~\cite{pinckney2025comprehensiveverilogdesignproblems}           & 4.93  & 0.12  \\
e7 & rgb\_color\_space\_conv~\cite{pinckney2025comprehensiveverilogdesignproblems} & 6.49 & 0.06 \\
e8 & spi\_complex\_mult~\cite{pinckney2025comprehensiveverilogdesignproblems}           & 2.01  & 0.11 \\
e9 & pooling~\cite{utlca_pooling}                                     & 2.04  & 0.08  \\
e10 & cryptech\_uart~\cite{secworks_uart}            & 2.70  & 0.06   \\
e11 & float\_adder~\cite{samidhm_float}                                & 3.36 & 0.10  \\
e12 & sha1\_top~\cite{secworks_sha1}                 & 3.81  & 0.04  \\
e13 & chacha\_top~\cite{secworks_chacha}             & 3.17  & N/A  \\
h1 & sd\_controller\_top~\cite{mczerski_sdcard}     & 0.45  & N/A  \\
h2 & trng\_top~\cite{secworks_trng}                 & 1.07 & N/A   \\
h3 & can~\cite{can_design}                 & 2.05 & N/A\\
h4 & axi\_crossbar~\cite{axi_crossbar}       & 0.44 & N/A \\
h5 & jpeg\_core~\cite{jpeg_core}     & 1.86  & N/A\\
h6 & ethernet\_mac~\cite{eth_mac}     & 1.02  & N/A \\
\hline
\end{tabular}
\label{tab:ratio-metrics}
\vspace{-4mm}
\end{table}

 \subsection{LLM Size Sensitivity} \label{sec:llm_size_sensitivity}

\begin{figure}[t]
    \centering
    \includegraphics[width=\linewidth]{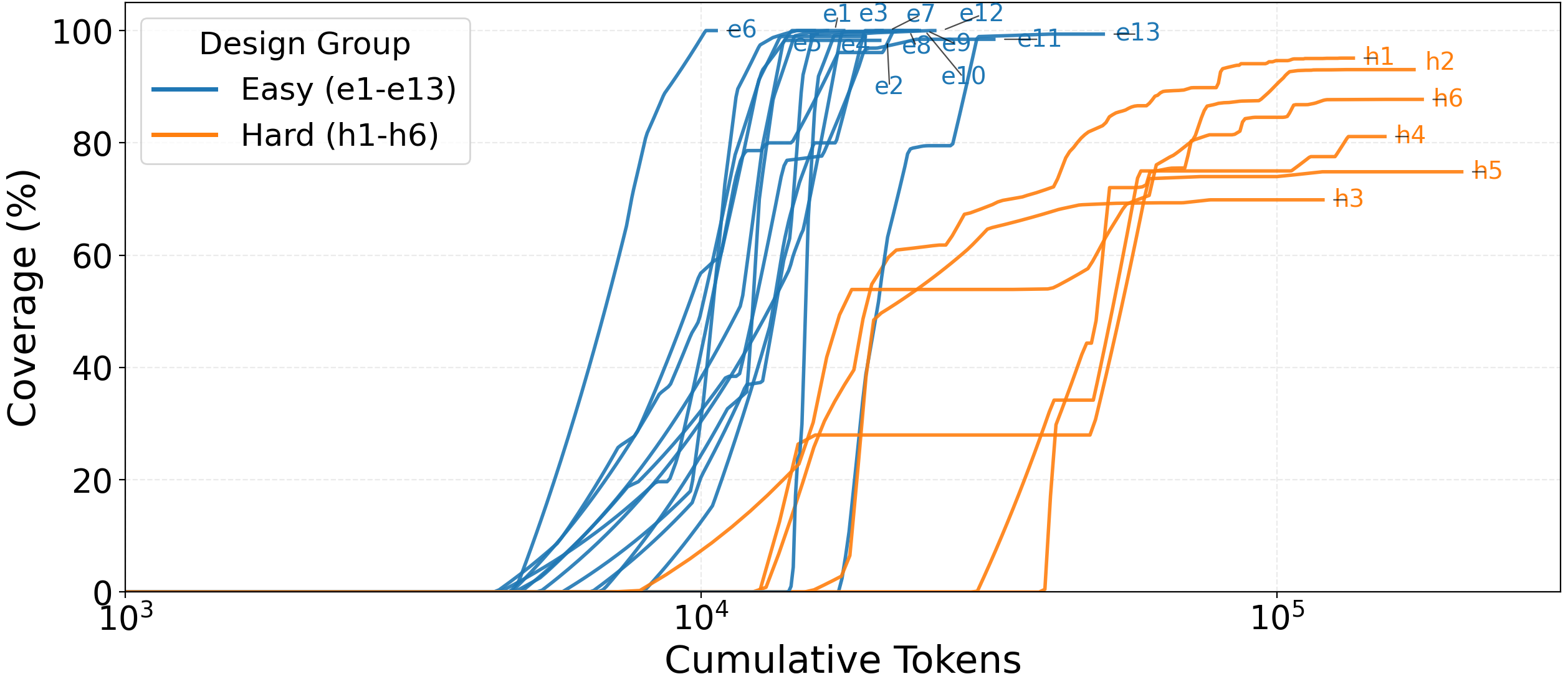}
    \caption{Coverage achieved vs. tokens consumed for GPT-5-mini}
    \label{fig:coverage_curves_mini}
    \vspace{-3mm}
\end{figure}

\begin{figure}[t]
    \centering
    \includegraphics[width=\linewidth]{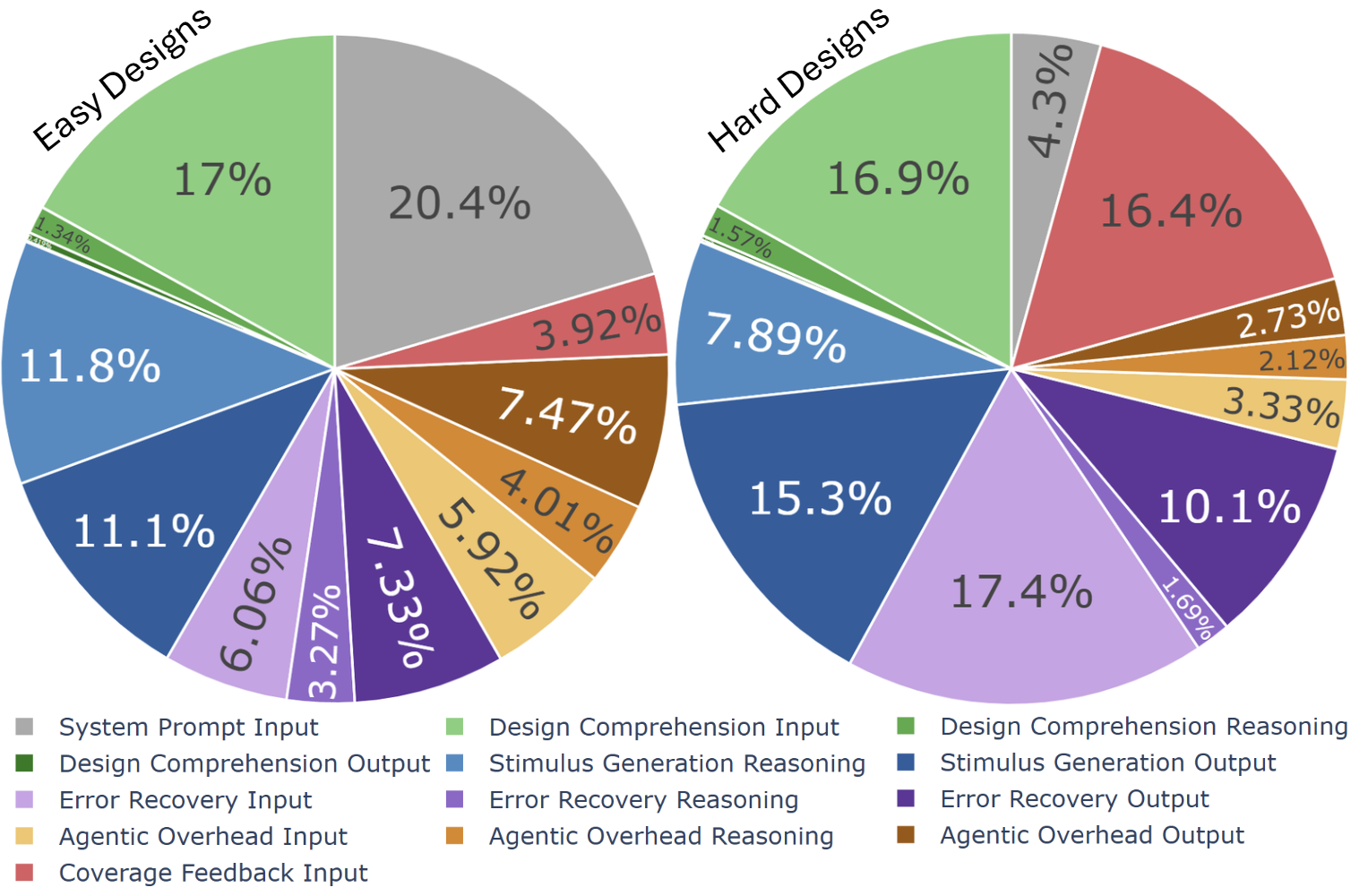}
    \caption{Average token allocation across designs  with GPT-5-mini.}
    \label{fig:token_alloc_mini}
    \vspace{-5mm}
\end{figure}

We evaluate our enhanced framework on a smaller/cheaper model - GPT-5-mini.
Fig. \ref{fig:coverage_curves_mini} shows the coverage achieved across all designs.
On easy designs, GPT-5-mini achieves 99.6\% coverage on average, requiring ${\sim}$1.7$\times$ more tokens than GPT-5.2.
On hard designs, however, the coverage generally degrades with complexity.
\textit{SD-card}: $-$2.9pp. 
\textit{trng}: $-$5.4pp.
\textit{ethmac}: $-$11.3pp,
\textit{AXI crossbar}: $-$14.9pp and 
\textit{CAN}: $-$29.9pp.
To understand further, we plot the token allocation breakdown.
Fig \ref{fig:token_alloc} shows that GPT-5.2 allocates 28.1\% of hard-design tokens to Stimulus Generation reasoning, whereas Fig \ref{fig:token_alloc_mini} shows GPT-5-mini allocates only 7.9\% (3.5$\times$ less). Conversely, GPT-5-mini's code output is 15.3\% vs.\ GPT-5.2's 17.3\%, implying that mini writes the same proportion of code with less reasoning, producing lower-quality stimulus that fails to close R1/R2 holes. 

\subsection{Cost-Coverage Tradeoff}

\begin{figure}[t]
    \centering
    \includegraphics[width=\linewidth]{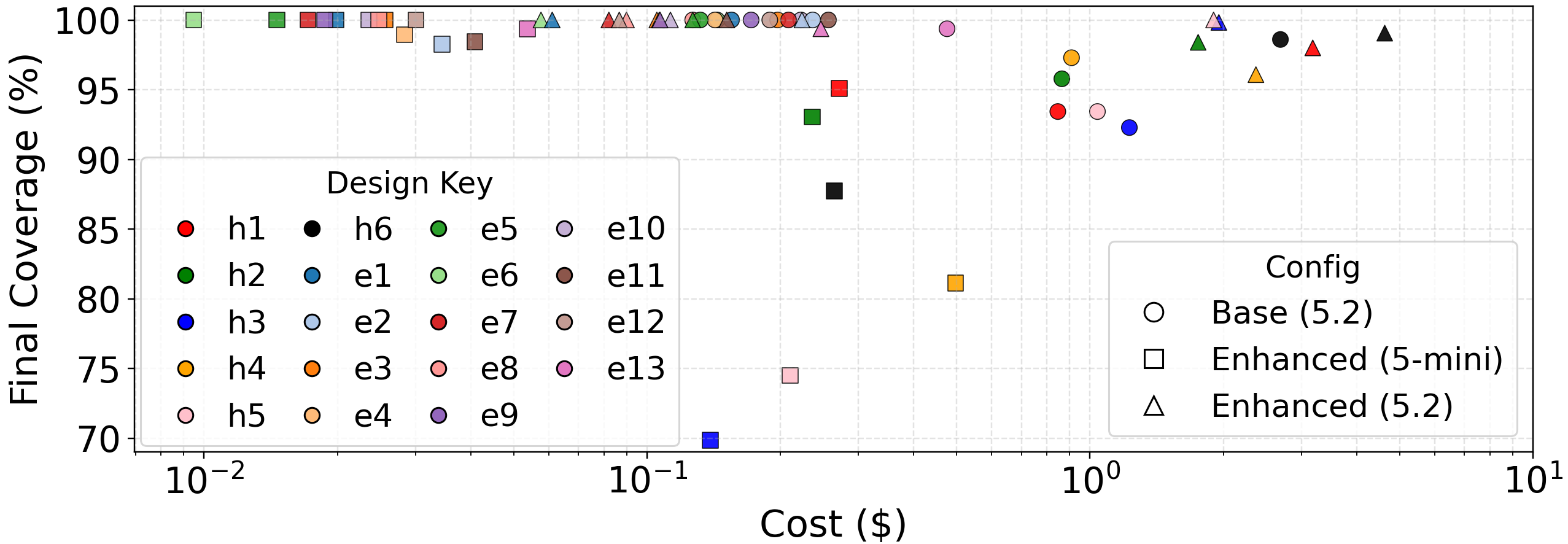}
    \caption{Cost vs. coverage tradeoff for various execution configs}
    \label{fig:cost-cov-tradeoff}
    \vspace{-6mm}
\end{figure}

Smaller models are cheaper but struggle to achieve coverage, while generic frameworks use more tokens and domain-specialized agents reduce them.
We evaluate whether domain specialization compensates for coverage loss from smaller models.
We evaluate the cost vs. coverage tradeoff in
Fig. \ref{fig:cost-cov-tradeoff}. It plots the coverage achieved and the dollar cost of tokens consumed across the three experiment configurations - Base agent with GPT 5.2, Enhanced agent with GPT 5.2, and Enhanced agent with GPT 5-mini.

Across nearly all designs, the enhanced agent (GPT-5.2 and GPT-5-mini) achieves comparable or higher coverage than the base configuration at significantly lower cost. The base agent remains costly regardless of design complexity due to infrastructure overhead (Section \ref{sec:proposal}). In contrast, the enhanced agent cuts cost by an order of magnitude, indicating that domain-specialized workflows eliminate non-productive token expenditure.

The enhanced GPT-5-mini configuration occupies the lowest-cost region of the plot while achieving near 95–100\% coverage on majority of designs. This demonstrates that domain specialization can partially compensate for reduced model capability.
However, this trend breaks down for more complex designs, where GPT-5-mini exhibits coverage degradation (e.g., in the 70–90\% range), despite comparable or slightly higher cost. This aligns with the reasoning-frontier limitations identified earlier where insufficient reasoning capacity cannot be offset by better tooling alone.


The enhanced GPT-5.2 configuration consistently achieves the highest coverage (97–100\%) across all designs while maintaining moderate cost. Importantly, it avoids the low-coverage outliers observed with GPT-5-mini, suggesting that higher-capacity models are necessary to reliably close “last-mile” coverage holes. These results suggest that a hybrid deployment strategy could be useful:
For easy-to-moderate designs, domain-specialized pipelines paired with smaller models offer the best cost-efficiency.
For complex designs, smaller models can be used for initial coverage ramp-up, but escalation to larger models is required once coverage plateaus.


\section{Conclusion} \label{sec:conclusion}

We presented CovAgent, a two-tiered open-source agentic framework for hardware verification coverage closure, enabling systematic study of inference-time token allocation and coverage hole classification. Across 19 designs, our analysis yields three key findings. 
First, the token allocation analysis enabled by our framework shows that only 12–18\% of inference-time tokens produce actual testbench code, while design comprehension reasoning scales disproportionately with design size. 
Second, the agent-generated human-audited coverage hole taxonomy reveals that residual coverage holes are split between methodology-bound ceilings (tied-off hardware, infeasible boundaries, dead code) and reasoning frontiers (protocol sequencing, pipeline warm-up, timing precision). 
Third, domain-specific workflow (annotated source feedback, structured tools, cumulative coverage merging) yields 4–13$\times$ token reduction at coverage parity compared to a general-purpose baseline.
By exposing both the limits and opportunities of agentic verification, CovAgent provides empirical insights for building more efficient and practically deployable verification systems.




\bibliographystyle{ACM-Reference-Format}
\bibliography{refs.bib}

\end{document}